\documentclass[traditabstract]{aa} 
\usepackage{graphicx,natbib,longtable,txfonts,lscape}
\usepackage[switch]{lineno}

\newcommand{\FRo}{FR0{\sl{CAT}}}

\newcommand{\WHz}{\>{\rm W}\,{\rm Hz}^{-1}}

\usepackage{color}

\begin{document}

\title{The low-frequency properties of FR~0 radio galaxies}
\author{A. Capetti\inst{1} \and R.~D. Baldi\inst{2} \and
  M. Brienza\inst{3,4} \and R. Morganti\inst{5,6} \and  G. Giovannini\inst{3,4}}

\institute{INAF - Osservatorio Astrofisico di
  Torino, Strada Osservatorio 20, I-10025 Pino Torinese, Italy
  \and School of Physics and Astronomy, University of
  Southampton, Southampton, SO17 1BJ, UK \and Dipartimento di
  Fisica e Astronomia, Universit\`a di Bologna, Via P. Gobetti 93/2,
  I-40129, Bologna, Italy \and INAF - Istituto di Radio
  Astronomia, Via P. Gobetti 101, I-40129 Bologna, Italy \and
  ASTRON, the Netherlands Institute of Radio Astronomy, Postbus 2,
  NL-7990 AA, Dwingeloo, the Netherlands \and Kapteyn
  Astronomical Institute, University of Groningen, PO Box 800, NL-9700
  AV Groningen, the Netherlands} \date{}

\abstract{Using the Alternative Data Release of the TIFR GMRT Sky
  Survey (TGSS), we studied the low-frequency properties of FR 0 radio
  galaxies, the large population of compact radio sources associated
  with red massive early-type galaxies revealed by surveys at 1.4 GHz.
  We considered TGSS observations from FR0{\sl CAT}, a sample formed
  by 104 FR~0s at $z<0.05$: all but one of them are covered by the
  TGSS, and 43 of them are detected above a 5$\sigma$ limit of 17.5
  mJy. No extended emission has been detected around the FR~0s,
  corresponding to a luminosity limit of $\lesssim 4 \times 10^{23}$ W
  Hz$^{-1}$ over an area of 100 kpc $\times$ 100 kpc. All but eight
  FR~0s have a flat or inverted spectral shape ($\alpha < 0.5$)
  between 150 MHz and 1.4 GHz: this spectral behavior confirms the
  general paucity of optically thin extended emission within the TGSS
  beam, as is expected for their compact 1.4 GHz morphology.

  Data at 5 GHz were used to build their radio spectra, which are
  also generally flat at higher frequencies. By focusing on a
  sub-sample of FR~0s with flux density $>$ 50 mJy at 1.4 GHz, we found
  that $\sim$75\% of them have a convex spectrum, but a smaller
  curvature than the more powerful gigahertz peaked-spectrum sources
  (GPS). The typical FR~0s radio spectrum is better described by a
  gradual steepening toward high frequencies, rather than to a
transition from an optically-thick to an optically-thin regime,
possibly observed in only $\sim 15\%$ of the sample.}

\maketitle

\section{Introduction}
\label{intro}

The identification of the optical counterparts of the large number of
sources detected in recent radio surveys at 1.4 GHz (e.g.,
\citealt{best12}) revealed that the majority of the radio sources
associated with low redshift galaxies show compact emission, with
sizes $\lesssim$ 10 kpc \citep{baldi09}. This population of compact
objects was poorly represented in earlier surveys (performed at lower
frequency and a higher flux density threshold), which were instead
dominated by sources extending over scales of hundreds of kpc (e.g.,
\citealt{hardcastle98}). Due to the lack of extended radio
  emission, these ``compact'' sources were named ``FR~0s''
\citep{ghisellini11,sadler14,baldi15}, a convenient way to include
them into the canonical \citet{fanaroff74} classification scheme of
radio galaxies (RGs). The information available from observations
  of FR~0s is generally very limited, even in the radio band. As a
consequence, it is still unclear what the nature of these sources is,
and how they are related to the other classes of RGs.

\citet{baldi18} selected a sample of compact radio sources named
\FRo, in order to perform a systematic study of FR~0s. \FRo\ consists
of 104 compact radio sources with redshift $\leq 0.05$ selected by
combining observations from the National Radio Astronomy Observatory
Very Large Array Sky Survey (NVSS; \citealt{condon98}), the Faint
Images of the Radio Sky at Twenty centimeters survey, (FIRST,
\citealt{becker95,helfand15}), and Sloan Digital Sky Survey (SDSS;
\citealt{york00b}).  In the catalog, \citeauthor{baldi18} included the
sources brighter than 5 mJy and with a limit to the deconvolved
angular size of 4$\arcsec$ in the FIRST images, corresponding to a
linear size $\lesssim$ 5 kpc. Their radio luminosities at 1.4 GHz are
in the $10^{22} \lesssim L_{1.4} \lesssim 10^{24} \WHz$range.

\citet{baldi19} obtained high-resolution multi-frequency radio images
of a sub-sample of 18 FR~0s randomly extracted from \FRo. Although the
observations reach an angular resolution of $\sim$0\farcs3
(corresponding to $\sim 250$ pc at the median redshift of these
  sources, $z=$0.04), 14 of the FR~0s are still unresolved, while
the remaining four extend over only a few kpc. These observations
confirm the general lack of extended radio emission and the high core
dominance of FR~0s when compared to the FR~Is of the 3C sample: 
  in FR~Is the fraction of nuclear emission is typically $10^{-2}$
  \citep{baldi19}.

The origin of the different nature of FR~0s with respect to the
extended RGs still remains to be understood.  While the appearance in
the radio images of FR~0s and FR~Is is radically different, the
nuclear and host galaxies' properties of these two classes are very
similar \citep{baldi18,torresi18}. A scenario in which FR~0s are young
RGs that will all eventually evolve into extended radio sources cannot
be reconciled with the large space density of FR~0s, five times more
abundant than FR~Is. FR~0s might instead be recurrent sources,
characterized by short phases of activity. \citet{baldi15} suggested
that the jet properties of FR~0s might be intrinsically different from
those of the FR~Is, for example, the former class with lower-bulk Lorentz
factors.

In this framework, low-frequency radio observations of FR~0s might
play an important role, as they can be used to address the following
questions: (1) Do FR~0s show low-frequency extended emission?
Compactness is the main defining characteristic of FR~0s and one
possibility to account for this property is that they are recurrent
sources. In this case, we could expect to detect remnant emission from
a previous cycle of activity. This is typically characterized by a
very steep spectrum, and is therefore best observable at
MHz-frequencies.  (2) What is the low-frequency spectral shape of
FR~0s? Observations at high resolution, required to spatially isolate
any small-scale extended emission, are only available for a minority
of FR~0s. The spectral index information can be used to infer the
fraction of optically thin, hence extended, emission present in FR~0s
overcoming the limited spatial resolution.  (3) What is the fraction
of young sources among the FR~0s? Such objects can be found by looking
for the characteristic signature of young radio galaxies, meaning,
their low-frequency spectral cut-off due to either synchrotron
self-absorption \citep{kellermann66,hodges84} or free-free absorption
\citep{kellermann66,bicknell97}.

The 150 MHz continuum survey performed with the Giant Metrewave Radio
Telescope (GMRT; \citealt{swarup91,intema17}) named TGSS (Tata
Institute of Fundamental Research GMRT Sky Survey) offers the first
possibility to gather low-frequency data with the combination of
sensitivity and spatial resolution required for the study of FR~0s.
In particular, we will focus on the TGSS observations of the
\FRo\ sample.

The paper is organized as follows. In Sect. \ref{sample}, we list the
data available from the TGSS observations of the \FRo\ sources. In
Sect.~\ref{results}, we present our results, which are then discussed in
Sect.~\ref{discussion}. In Sect.~\ref{summary}, we summarize the
results and draw our conclusions.

\begin{table*}
\caption{Radio properties of the sample}
\begin{tabular}{c r r r r r | c r r r r r}
\hline
SDSS~name &  {\small F(150)} &{\small  F(1.4) } & {\small  F(5)}  & $\alpha_1$ & $\alpha_2$ &  SDSS~name & {\small F(150)} &{\small  F(1.4)} & {\small F(5)}  &  $\alpha_1$ & $\alpha_2$  \\
\hline
010852.48$-$003919.4 &  ---      &  10.9  &   --- &$<$ 0.22 &    ---   & 123011.85+470022.7    & 128.8   &  93.8  &      73     &     0.14 &     0.20\\ %001    \\ %053 
011204.61$-$001442.4 &  ---      &  17.9  &   --- &$<$ 0.00 &    ---   & 124318.73+033300.6    & 304.0   &  63.5  &      $<$30  &     0.70 & $>$ 0.60\\ %002    \\ %054 
011515.78+001248.4   &  ---      &  42.6  & $<$33 &$<$-0.40 &$>$ 0.21  & 124633.75+115347.8    & 36.1    &  61.2  &      40     &    -0.24 &     0.34\\ %003    \\ %055 
015127.10$-$083019.3 &  60.8     &  35.7  &   --- &    0.23 &    ---   & 125027.42+001345.6    & ---     &  54.5  &      82     & $<$-0.51 &    -0.33\\ %004    \\ %056 
020835.81$-$083754.8 &  ---      &  28.4  &   --- &$<$-0.20 &    ---& 125409.12$-$011527.1  & ---     &   7.7  &        ---  & $<$ 0.37 &     --- \\ %005    \\ %057 
075354.98+130916.5   &  ---      &   7.4  & $<$23 &$<$ 0.39 &$>$-0.92  & 130404.99+075428.4    & ---     &  10.5  &      $<$26  & $<$ 0.23 & $>$-0.74\\ %006    \\ %058 
080716.58+145703.3   &  43.2     &  28.4  & 28*   &    0.19 &    0.01  & 130837.91+434415.1    & 107.4   &  58.4  &      47     &     0.27 &     0.17\\ %007    \\ %059 
083158.49+562052.3   &  26.1$^a$ &   9.0  & $<$18 &    0.48 &$>$-0.56  & 133042.51+323249.0    & ---     &  17.9  &      $<$19  & $<$-0.01 & $>$-0.04\\ %008    \\ %060 
083511.98+051829.2   &  ---      &  10.1  & $<$29 &$<$ 0.24 &$>$-0.84  & 133455.94+134431.7    & 41.3    &  39.4  &      $<$23  &     0.02 & $>$ 0.43\\ %009    \\ %061 
084102.73+595610.5   &  ---      &   8.9  & $<$18 &$<$ 0.30 &$>$-0.57  & 133621.18+031951.0    & ---     &  30.4  &      $<$30  & $<$-0.25 & $>$ 0.01\\ %010    \\ %062 
084701.88+100106.6   &  22.4$^a$ &  23.7  & $<$25 &   -0.03 &$>$-0.04  & 133737.49+155820.0    & 41.6    &  26.9  &      $<$22  &     0.20 & $>$ 0.16\\ %011    \\ %063 
090652.79+412429.7   &  ---      &  51.8  & 65    &$<$-0.49 &   -0.18  & 134159.72+294653.5    & ---     &  10.4  &      $<$19  & $<$ 0.23 & $>$-0.49\\ %012    \\ %064 
090734.91+325722.9   & {\small no data}  &  46.9  & $<$19 &    ---  &$>$-0.76  & 135036.01+334217.3    & 302.3   & 101.3  &      79     &     0.49 &     0.20\\ %013    \\ %065 
090937.44+192808.2   &  168.2    &  69.1  & 106   &    0.40 &   -0.34  & 135226.71+140528.5    & ---     &  25.5  &      $<$23  & $<$-0.17 & $>$ 0.09\\ %014    \\ %066 
091039.92+184147.6   &  168.8    &  50.0  & 47*   &    0.54 &    0.05  & 140528.32+304602.0    & ---     &   7.4  &      $<$19  & $<$ 0.39 & $>$-0.76\\ %015    \\ %067 
091601.78+173523.3   &  94.8     &  24.5  & $<$21 &    0.61 &$>$ 0.12  & 141451.35+030751.2    & ---     &  26.7  &      85     & $<$-0.19 &    -0.93\\ %016    \\ %068 
091754.25+133145.5   &  16.8     &  22.9  & $<$23 &   -0.14 &$>$-0.01  & 141517.98$-$022641.0  & ---     &  18.9  &        ---  &          & $>$-0.02\\ %017    \\ %069 
093003.56+341325.3   &  81.5     &  33.1  & $<$19 &    0.40 &$>$ 0.46  & 142724.23+372817.0    & ---     &  10.8$^b$ &   $<$18  & $<$ 0.22 & $>$-0.42\\ %018    \\ %070 
093346.08+100909.0   &  99.7     &  56.6  & 80    &    0.25 &   -0.28  & 143156.59+164615.4    & ---     &   8.7  &      $<$21  & $<$ 0.31 & $>$-0.73\\ %019    \\ %071 
093938.62+385358.6   &  ---      &   6.1  & $<$18 &$<$ 0.47 &$>$-0.88  & 143312.96+525747.3    & ---     &  15.6  &      $<$18  & $<$ 0.05 & $>$-0.12\\ %020    \\ %072 
094319.15+361452.1   &  35.4     &  75.1  & 81    &   -0.34 &   -0.06  & 143424.79+024756.2    & ---     &   7.3  &      $<$31  & $<$ 0.39 & $>$-1.15\\ %021    \\ %073 
100549.83+003800.0   &  ---      &  24.1  & $<$32 &$<$-0.14 &$>$-0.24  & 143620.38+051951.5    & ---     &  18.7  &      $<$29  & $<$-0.03 & $>$-0.34\\ %022    \\ %074 
101329.65+075415.6   &  ---      &   7.8  & $<$26 &$<$ 0.36 &$>$-0.98  & 144745.52+132032.2    & ---     &   6.7  &      $<$23  & $<$ 0.43 & $>$-1.00\\ %023    \\ %075 
101806.67+000559.7   &  ---      &  14.3  & $<$33 &$<$ 0.09 &$>$-0.67  & 145216.49+121711.5    & ---     &   8.0  &      $<$24  & $<$ 0.35 & $>$-0.87\\ %024    \\ %076 
102403.28+420629.8   &  ---      &   6.0  & $<$18 &$<$ 0.48 &$>$-0.88  & 145243.25+165413.4    & 17.9$^a$&  17.5  &      $<$21  &     0.01 & $>$-0.17\\ %025    \\ %077 
102511.50+171519.9   &  ---      &  10.2  & $<$21 &$<$ 0.24 &$>$-0.59  & 145616.20+203120.6    & 68.3    &  25.8  &      $<$20  &     0.44 & $>$ 0.21\\ %026    \\ %078 
102544.22+102230.4   &  90.4     &  76.7  & 60    &    0.07 &    0.20  & 150152.30+174228.2    & 18.6$^a$&  18.6  &      $<$21  &     0.14 & $>$-0.10\\ %027    \\ %079 
103719.33+433515.3   &  260.9    & 132.2  & 66    &    0.30 &    0.56  & 150425.68+074929.7    & ---     &   7.8  &      $<$27  & $<$ 0.36 & $>$-0.99\\ %028    \\ %080 
103952.47+205049.3   &  ---      &   6.9  & $<$20 &$<$ 0.42 &$>$-0.85  & 150601.89+084723.2    & ---     &   8.3  &      $<$26  & $<$ 0.33 & $>$-0.91\\ %029    \\ %081 
104028.37+091057.1   &  128.5    &  68.5  & 31    &    0.28 &    0.64  & 150636.57+092618.3    & 27.8$^a$&  27.8  &      $<$25  &     0.03 & $>$ 0.08\\ %030    \\ %082 
104403.68+435412.0   &  23.4     &  32.4  & 28    &   -0.15 &    0.12  & 150808.25+265457.6    & ---     &  20.3  &      $<$19  & $<$-0.07 & $>$ 0.04\\ %031    \\ %083 
104811.90+045954.8   &  175.3    &  49.1  & $<$29 &    0.57 &$>$ 0.43  & 152010.94+254319.3    & ---     &  18.3  &      $<$19  & $<$-0.02 & $>$-0.05\\ %032    \\ %084 
104852.92+480314.8   &  63.7     &  19.2  & $<$18 &    0.54 &$>$ 0.05  & 152151.85+074231.7    & ---     &  11.7  &      $<$27  & $<$ 0.18 & $>$-0.66\\ %033    \\ %085 
105731.16+405646.1   &  33.2     &  44.8  & 31    &   -0.13 &    0.30  & 153016.15+270551.0    & ---     &  13.3  &      $<$19  & $<$ 0.12 & $>$-0.30\\ %034    \\ %086 
111113.18+284147.0   &  56.4     &  41.1  & 56*   &    0.14 &   -0.25  & 154147.28+453321.7    & ---     &   8.9  &      $<$18  & $<$ 0.30 & $>$-0.57\\ %035    \\ %087 
111622.70+291508.2   &  ---      &  71.5  & $<$19 &$<$-0.63 &$>$ 1.06  & 154426.93+470024.2    & 46.5    &  17.6  &      $<$18  &     0.43 & $>$-0.02\\ %036    \\ %088 
111700.10+323550.9   &  ---      &  17.6  & $<$19 &$<$-0.00 &$>$-0.05  & 154451.23+433050.6    & ---     &  11.5  &      $<$18  & $<$ 0.19 & $>$-0.36\\ %037    \\ %089 
112029.23+040742.1   &  ---      &   7.5  & $<$30 &$<$ 0.38 &$>$-1.10  & 155951.61+255626.3    & 67.7$^a$&  29.2$^b$ &   $<$19  &     0.38 & $>$ 0.33\\ %038    \\ %090 
112256.47+340641.3   &  46.5     &  16.6  & $<$19 &    0.46 &$>$-0.09  & 155953.99+444232.4    & 188.5   &  59.5  &      $<$18  &     0.52 & $>$ 0.96\\ %039    \\ %091 
112625.19+520503.5   &  ---      &   9.0  & $<$18 &$<$ 0.30 &$>$-0.56  & 160426.51+174431.1    & 45.7    &  96.0  &      143    &    -0.33 &    -0.32\\ %040    \\ %092 
112727.52+400409.4   &  43.0     &  13.8  & $<$18 &    0.51 &$>$-0.21  & 160523.84+143851.6    & ---     &   8.6  &      $<$23  & $<$ 0.32 & $>$-0.78\\ %041    \\ %093 
113449.29+490439.4   &  107.0    &  33.0  & $<$18 &    0.53 &$>$ 0.49  & 160641.83+084436.8    & ---     &   9.3  &      $<$26  & $<$ 0.28 & $>$-0.82\\ %042    \\ %094 
113637.14+510008.5   &  ---      &   9.0  & $<$18 &$<$ 0.30 &$>$-0.56  & 161238.84+293836.9    & ---     &  27.4  &      $<$19  & $<$-0.20 & $>$ 0.29\\ %043    \\ %095 
114230.94$-$021505.3 &  ---      &   8.8  &   --- &$<$ 0.31 &    ---   & 161256.85+095201.5    & 52.6    &  21.7  &      $<$25  &     0.40 & $>$-0.11\\ %044    \\ %096 
114232.84+262919.9   &  ---      &  42.0  & 51    &$<$-0.39 &   -0.16  & 162146.06+254914.4    & ---     &   9.1  &      $<$19  & $<$ 0.29 & $>$-0.61\\ %045    \\ %097 
114804.60+372638.0   &  ---      &  29.1  & 24    &$<$-0.23 &    0.16  & 162846.13+252940.9    & ---     &  25.2  &      $<$19  & $<$-0.16 & $>$ 0.21\\ %046    \\ %098 
115531.39+545200.4   &  ---      &  31.2  & $<$18 &$<$-0.26 &$>$ 0.44  & 162944.98+404841.6    & ---     &   7.7  &      $<$18  & $<$ 0.37 & $>$-0.68\\ %047    \\ %099 
120551.46+203119.0   &  92.6     &  89.9  & 52    &    0.01 &    0.44  & 164925.86+360321.3    & ---     &  11.9  &      $<$18  & $<$ 0.17 & $>$-0.35\\ %048    \\ %100 
120607.81+400902.6   &  ---      &   9.5  & $<$18 &$<$ 0.27 &$>$-0.51  & 165830.05+252324.9    & ---     &  13.1  &      $<$19  & $<$ 0.13 & $>$-0.32\\ %049    \\ %101 
121329.27+504429.4   &  178.1    &  96.5  & $<$18 &    0.27 &$>$ 1.35  & 170358.49+241039.5    & 28.2    &  32.7  &      $<$20  &    -0.07 & $>$ 0.41\\ %050    \\ %102 
121951.65+282521.3   &  ---      &   8.7  & $<$19 &$<$ 0.31 &$>$-0.64  & 171522.97+572440.2    & 71.6    &  57.2  &      35     &     0.10 &     0.40\\ %051    \\ %103 
122421.31+600641.2   &  ---      &   6.1  & $<$18 &$<$ 0.47 &$>$-0.87  & 172215.41+304239.8    & 48.2    &   8.1  &      $<$19  &     0.80 & $>$-0.68\\ %052    \\ %104 
\hline
\end{tabular}
\label{tab1}
%\smallskip
\small{Column description: (1) name; (2 - 4) flux densities (in mJy)
  at 0.15, 1.4, and 5 GHz, respectively. (5 - 6) spectral indices
  between 0.15 and 1.4 GHz ($\alpha_1$) and 1.4 and 5 GHz
  ($\alpha_2$).  The sources marked with $^a$ are those not present in
  the TGSS catalog, and whose flux density was measured from the
  images, while those marked with ``---'' are not detected at 150 MHz
  with a threshold of 17.5 mJy. For the sources marked with $^b,$ we
  used the FIRST measurement instead of the NVSS one, due to the
  contamination of a nearby source. Sources outside the GB6 survey
  coverage are indicated with ``---'' in the last column, while the
  three FR~0s whose five GHz measurements are potentially contaminated by
  nearby sources are marked with an asterisk.}
\end{table*}

\begin{figure*}
\includegraphics[scale=1.00]{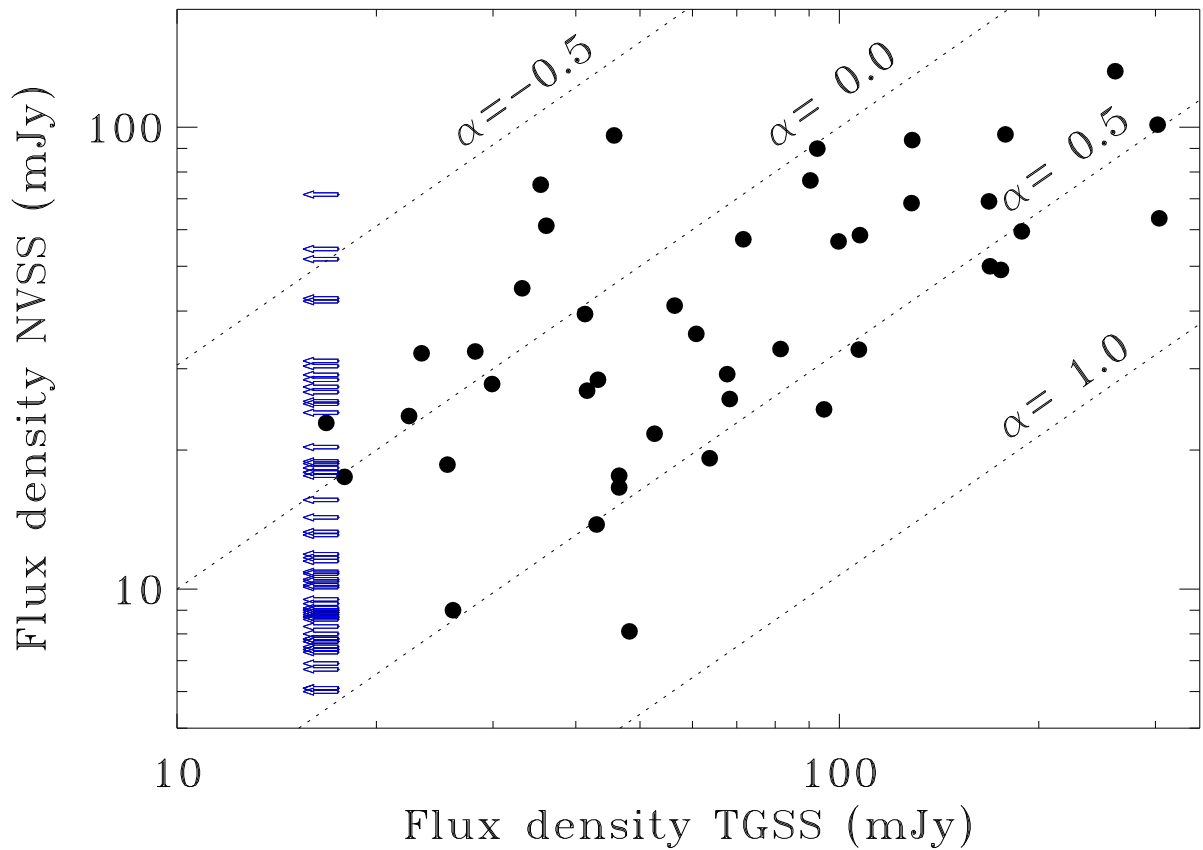}
\caption{Comparison of flux densities of FR0CAT sources at 150
  MHz and 1.4 GHz from TGSS and NVSS, respectively. The dotted
  lines represent loci of constant spectral indices $\alpha$ (defined
  as $F_{\nu}\propto\,\nu^{-\alpha}$) at the values indicated. The
  blue, left pointing arrows represent the upper limits in the TGSS.}
\label{tgss}
\end{figure*}

\section{The TGSS data}
\label{sample}

The low-frequency data for the FR~0s are taken from the TGSS.
\citet{intema17} produced a first alternative data release (TGSS ADR1)
obtained through independent re-processing of the TGSS data. The TGSS
ADR1 covers the declination range -53$^\circ$ $<$ $\delta$ $<$
+90$^\circ$ with images at 150 MHz, at a resolution of
$\sim$25\arcsec\ corresponding to a linear scale between 10 and 25 kpc
at the redshift of the sources considered. The median rms noise level
of the TGSS ADR1 images is 3.5 mJy beam$^{-1}$. The catalog contains
all sources above the 7$\sigma$ (24.5 mJy) threshold. The absolute
astrometry accuracy of the catalog is better than 2\arcsec.

All but one (namely J0907+35) of the 104 sources of the FR0{\sl CAT}
sample are covered by the TGSS. By adopting a search radius of
5\arcsec, we found an association for 37 out of the 104 \FRo\ sources
in the TGSS ADR1 catalog. However, the a priori knowledge of the
optical position of the FR~0s enables us to safely use a less
stringent limit, which we set at 5$\sigma$ (17.5 mJy). With this
strategy, we recover five additional FR~0 detections. The visual
inspection of the TGSS fields of all FR~0 sources also revealed an
association for J1559+25, not listed in the catalog, likely
because of the presence of a confusing bright (180 mJy at 1.4 GHz)
nearby (29\arcsec) object. We measured a TGSS flux density for
J1559+25 of 67.7 mJy. The total number of FR~0s with a 150 MHz flux
density measurement is then 43. In Table \ref{tab1}, we list the radio
data used for our analysis.

To study the spectral properties of the FR~0s, we also used data from
the Green Bank 6-cm survey (GB6, \citealt{gregory96}) and the
Westerbork Northern Sky Survey at 327 MHz (WENSS,
\citealt{rengelink97}). GB6 covers the Northern sky up to $\delta =
75^\circ,$ and it includes the position of all but seven
\FRo\ sources. The catalog threshold is generally 18 mJy, but it varies
with position within the survey area, being higher at lower
declinations (see Tab. \ref{tab1}). The WENSS covers the sky North of
$\delta = 30^\circ$, including 25 \FRo\ sources, with a limiting flux
density of 18 mJy, and a resolution of 54\arcsec$\times$54\arcsec
cosec$\delta$.

Due to the large beams of the TGSS, NVSS, WENSS, and GB6 (with
resolutions of 25\arcsec, 45\arcsec, 54\arcsec, and 3\arcmin,
respectively), there is the possibility of a contamination from
sources located at small distances from the FR0s. We inspected their
higher resolution FIRST images to address this issue. Concerning the
possible contamination of the first three surveys, we considered an
area with a radius of 45$\arcsec$ (the NVSS beam size) centered on the
targeted FR~0: we found that this includes a source present in the
FIRST catalog in only two cases: J1427+37 and J1559+25. J1427+37 is
not detected at either 150 MHz or 5 GHz. We already
  identified the confusing source in the TGSS image for J1559+25
(which is not visible in GB6). For these two objects, we used the
  FIRST flux density value at 1.4 GHz instead of the NVSS value.

To assess any possible contamination of the GB6 measurements, we only
considered the 23 FR~0s detected by this survey. For eight of them,
there is no FIRST source within 3$^\prime$. In 12 cases, the sources
within this radius have a flux density at 1.4 GHz between 3\% and
  10\% of the FR~0 considered: even in the case that these sources have an
  inverted spectrum, they do not produce a significant contribution
  ($>$25\%) of the measured 5 GHz flux density. In the remaining
three cases (namely, J0807+14, J0910+18, and J1111+28), there are
nearby sources with flux densities between 30\% and 90\% of the target
of interest. We marked them with an asterisk in Tab. \ref{tab1} to
highlight the possible contamination.

\begin{figure*}
\includegraphics[scale=0.45]{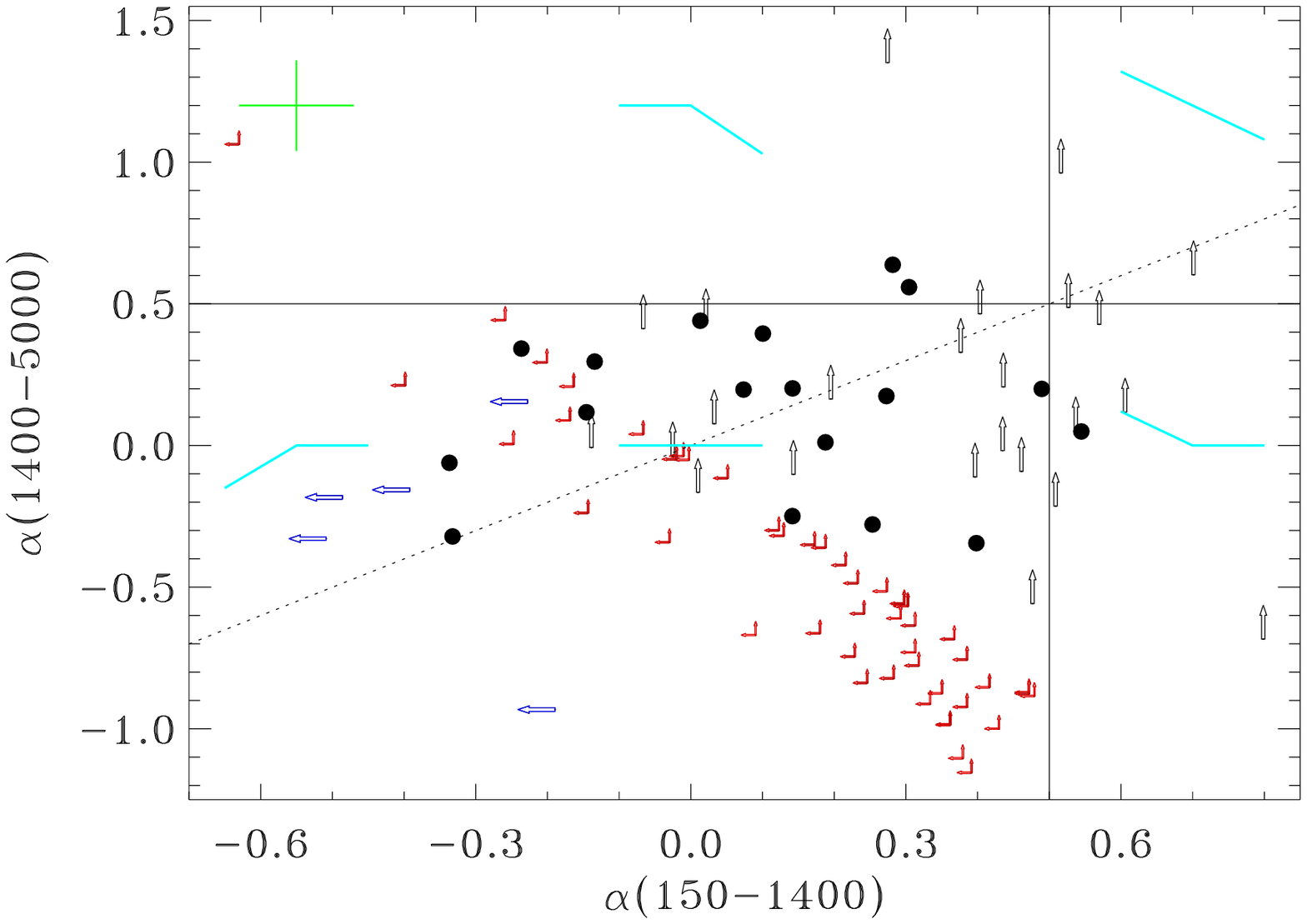}
\includegraphics[scale=0.45]{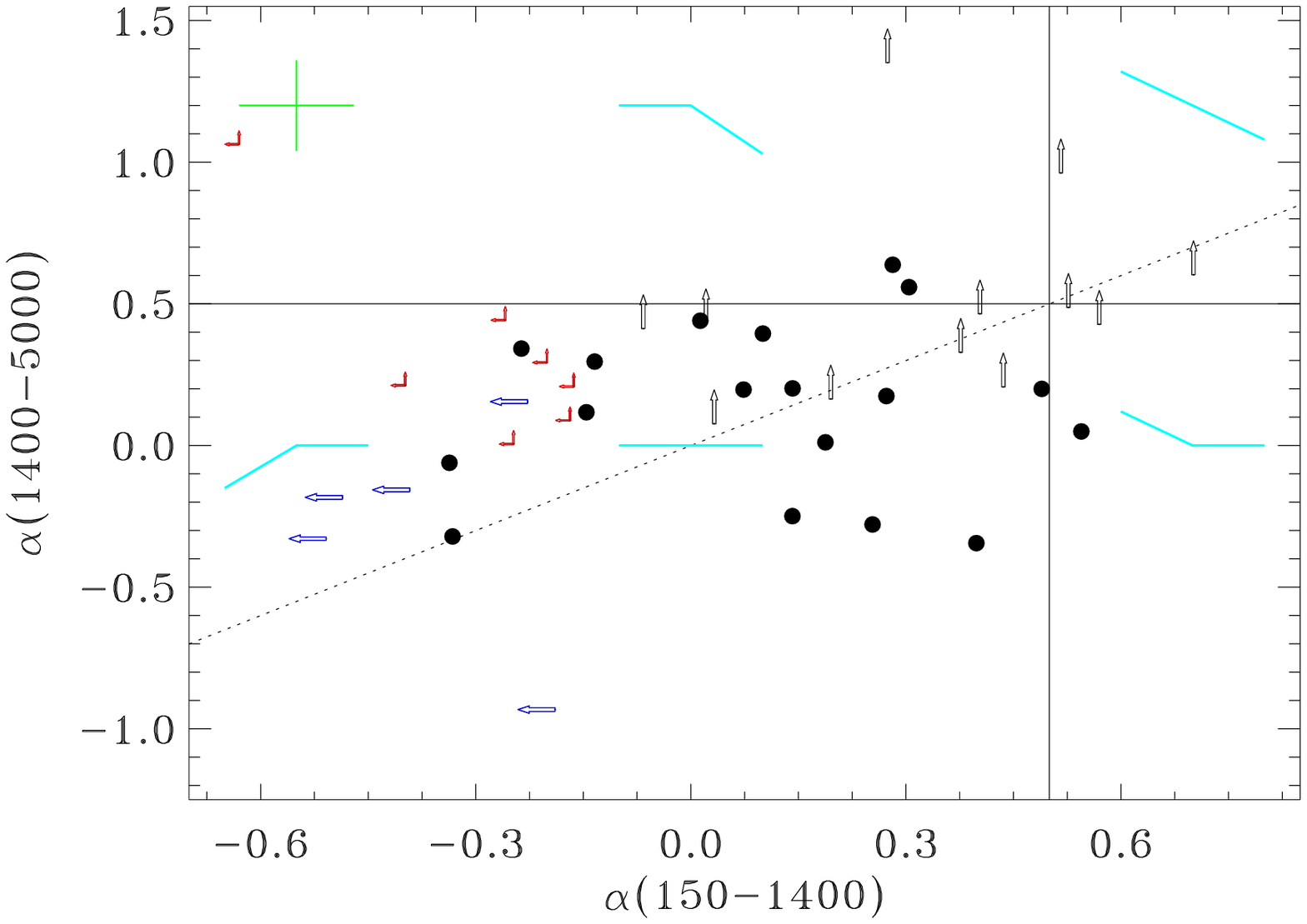}
\caption{Left: comparison of spectral indices measured between 150
  MHz and 1.4 GHz and 1.4 and 5 GHz. The double red arrows indicate
  objects only detected at 1.4 GHz, the blue arrows indicate those detected at
  1.4 and 5 GHz, but not at 150 MHz, the black arrows indicate those detected
  at 150 MHz and 1.4 GHz, but not at 5 GHz. The solid lines separate
  flat ($\alpha <0.5$) and steep sources. Sources located above the
  dotted line (marking an equal value of the two spectral indices) are
  objects with a convex spectrum. The green cross on the upper left
  corner represents the maximum error, corresponding to the sources
  with flux density equal to $5 \times \sigma$ in both TGSS and
  GB6. The cyan broken lines reproduce the typical spectral shape in
  different regions of the diagram.  Right: same as left panel but
  limited to 19 sources with $F_{\rm 1.4} > 50$ mJy.}
\label{three}
\end{figure*}

\section{Results}
\label{results}

The FR~0s detected in the TGSS images (with just one exception) show a
single unresolved radio component confirming the lack of large scale
extended emission characteristic of these sources seen in both the
FIRST and the NVSS. The only exception is J1521+07, also known as 3C~318.1,
located at the center of the MKW03S cluster: an extended radio source
with an extremely steep spectrum ($\alpha$=2.42 between 235 and 1280
MHz)\footnote{Spectral indices $\alpha$ are defined as
  $F_{\nu}\propto\,\nu^{-\alpha}$.} and a morphology dominated by a
peculiar concave arc-like structure (i.e. with its center of
  curvature located on the opposite side of the J1521+07) is seen
$\sim$ 40 kpc toward the south \citep{giacintucci07}, a structure
possibly not associated with the active nucleus.

We can estimate the limit on the flux density of any extended emission
around the FR~0s by considering a fiducial area of 100 kpc $\times$
100 kpc. At the median redshift of the \FRo\ sample of 0.037, this
corresponds to $\sim 135\arcsec \times 135\arcsec$. The distribution
of GMRT antenna baselines is such that the array at 150 MHz is
sensitive to extended emission on scales smaller than 68$^\prime$
\citep{intema17}, much larger than the angular sizes we are interested
in. We measured the flux density over $\sim 135\arcsec \times
135\arcsec$ in several regions of the FR~0 images. We found a median
value of five mJy with a dispersion of 39 mJy, corresponding to a
3$\sigma$ upper limit of $\sim$120 mJy.

In Fig. \ref{tgss}, we compare the flux densities of the FR0CAT sources
at 150 MHz and 1.4 GHz from the TGSS and NVSS, respectively. Since the
FR~0s are compact objects, we do not expect the different
resolution of the two surveys to significantly affect our
results. Conversely, variability is an important issue, particularly
because we are dealing with compact sources and considering that
$\sim$15 years separate the TGSS and NVSS observations. Therefore,
the results of this comparison for individual objects should be taken
with some caution.

The resulting spectral indices are steeper than $\alpha > 0.5$ only in
eight sources, meaning, the vast majority (92\%) of the FR~0s show a
flat ($0 < \alpha <0.5$) or inverted ($\alpha <0$) low-frequency
spectrum. This conclusion applies not only to the sources detected by
the TGSS, but also to those with a 150 MHz upper limit, because they
all correspond to slopes flatter than $\alpha < 0.47$. Several FR~0s
show an inverted spectrum: this is the case of eight of the TGSS
detected sources, to which we add the 21 undetected objects with an
upper limit $\alpha < 0$ to their spectral indices.  The fraction of
FR~0s with inverted spectrum is then at least 28\%, but it can be as
high as 66\%, depending on the slope of the remaining 39 FR~0s,
undetected in the TGSS, which are all consistent with a negative value
of $\alpha$.

\begin{figure*}
\includegraphics[scale=0.99]{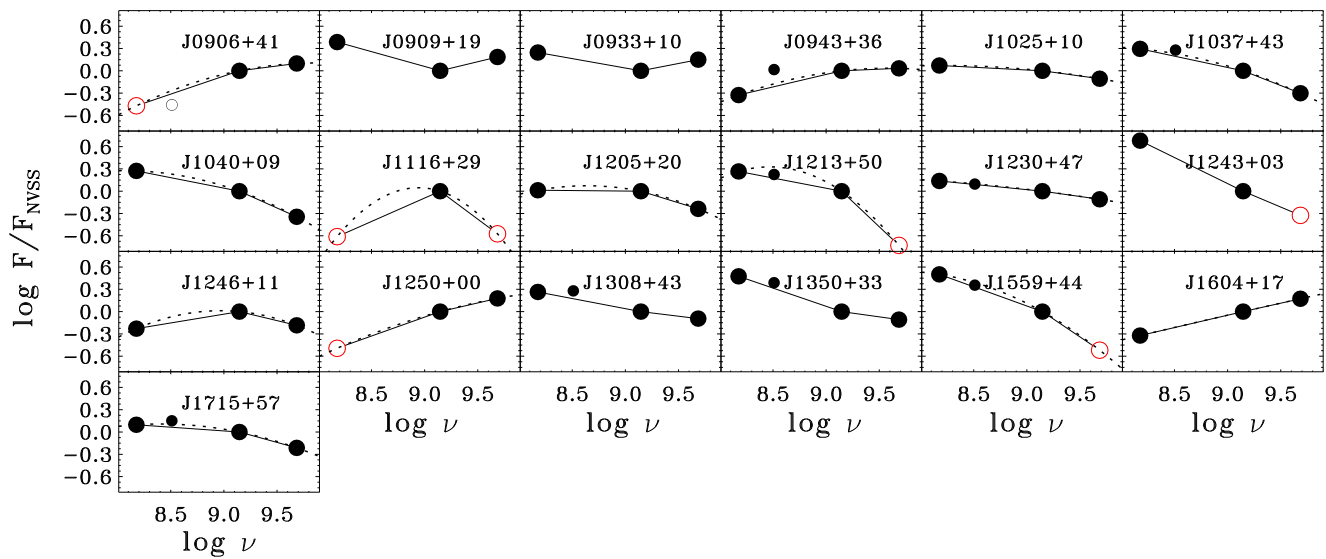}
\caption{Radio spectra of the 19 FR~0s with a 1.4 GHz flux
    density $> 50$ mJy, covered by both the TGSS and GB6. Flux
  densities are normalized to unity at 1.4 GHz. The empty red symbols
  correspond to upper limits. The dashed lines are the log-parabolae
  defined by the measurements at 150 MHz, 1.4 and 5 GHz which are
  shown for the sources with a convex spectrum. The smaller symbols
  are the available WENSS measurements at 327 MHz.}
\label{sed}
\end{figure*}

\begin{figure}
\includegraphics[scale=0.49]{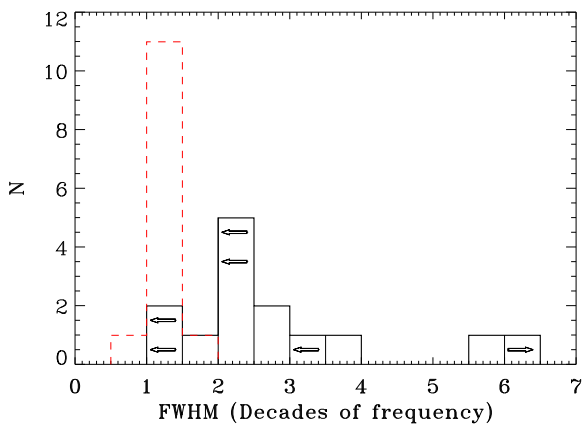}
\caption{Black histogram: distribution of FWHM (in decades of
  frequency) of 14 FR~0s with a convex radio spectrum extracted
  from 19 sources of bright sub-sample with $F_{1.4} > 50$
  mJy. The left pointing arrows correspond to upper limits of the FWHM
  due to the non detection at least at one frequency. The right
  pointing arrow represents the FWHM of J1604+17 whose measured value
  (12.5) exceeds the plot limit. The dashed red histogram reports the
  FWHM measured by \citet{odea91} in a sample of 13 GPS.}
\label{fwhist}
\end{figure}

To further investigate the spectral properties of the FR~0s, we also include the GB6 observations at 5 GHz in the analysis and compare the
spectral indices measured between 150 MHz and 1.4 GHz, and between 1.4
and 5 GHz (see Fig. \ref{three}). This comparison confirms the main conclusion on the paucity of FR~0s with a steep spectrum. The
GB6 measurements are also important to finding and to exploring the nature
of the sources with a convex spectrum. All sources with
$\alpha(150-1400) < \alpha(1400-5000) $, meaning, those located above the
dotted line (marking an equal value of the two spectral indices
considered) are objects with a convex spectrum.

A significant limitation of our analysis is related to the relatively
small fraction of objects detected at 150 MHz and (or) 5 GHz, due to the
higher flux density thresholds of the TGSS and GB6 data with respect
to the NVSS. We therefore preferred to restrict this analysis to the
sub-sample formed by the 19 FR~0s with a 1.4 GHz flux density $F_{\rm
  1.4} > 50$ mJy, covered by both the TGSS and GB6 area (see
Fig. \ref{three}, right panel). At least 14 of these sources have a
convex spectrum (see Fig. \ref{sed}).

Following the approach of \citet{odea91}, we fit the radio spectra
with a log-parabola and measured its full width at half maximum
(FWHM). We limited our study to the 14 sources with a convex spectrum and
  measured values typically between 1.5 and six decades (see
  Fig. \ref{fwhist}). The FWHM in five sources must be considered as
  an upper limit, because the source is not detected in at least one
  band. The median FWHM is between 2.2 and 2.4 decades, depending on
  the actual values of the upper limits.

The sampling of the radio spectra of FR~0s is clearly rather
limited. For seven of these sources, the spectral coverage can be
improved by including the measurements from the WENSS: overall, these
support the interpretation that the FR~0s' spectra are rather shallow.
In particular, in two of the sources not detected by the GB6 survey
(namely, J1213+50 and J1559+44), the fitted parabolas are consistent
with the WENSS data points, suggesting that the upper bounds to their
curvature do not differ significantly from their actual values.

\section{Discussion}
\label{discussion}

\subsection{Extended radio emission in FR~0s }

The TGSS images confirm the lack of spatially extended emission around
FR~0s, the main defining property of this class of sources. An
extended steep spectrum component might have been expected in two
cases. Firstly, FR~0s might be less efficient in the acceleration of the
relativistic electrons with respect to standard extended radio-galaxies, and
produce intrinsically steep radio spectra. Secondly, FR~0s might be
recurrent sources leaving behind a relic emission from previous
activity phases, characterized by a steep spectrum due to spectral
ageing. The limit to the luminosity of extended structures is $\sim 4
\times 10^{23}$ W Hz$^{-1}$ over an area of 100 kpc $\times$ 100 kpc
at the median distance of the FR 0 sample. As reference, the
edge-darkened FR~I sources selected from the same catalog of RGs from
which we extracted the FR~0s \citep{capetti17} have sizes between 60
and 120 kpc, and predicted luminosities at 150 MHz (when assuming a
spectral slope of 0.7 between 150 MHz and 1.4 GHz) in the $\sim
10^{24} - 10^{26}$ W Hz$^{-1}$range.

Nonetheless, very low surface brightness emission ($\sim$ 0.15 mJy
beam$^{-1}$) extending over $\sim 10$ kpc have been detected in
NGC~3998 \citep{frank16}. NGC~3998 is a nearby (z=0.0035) flat-spectrum radio source, unresolved in the FIRST images, fulfilling all
requirements for an FR~0 classification.\footnote{NGC~3998 is not
  included in the \FRo,\ because the SDSS did not obtain its optical
  spectrum.}  This individual example indicates that, although FR~0s
and FR~Is are different classes of sources, deeper observations are
needed to explore their relationship in greater depth, and in
particular, to establish whether FR~0s are able to produce large-scale
jets.

\subsection{Radio spectral properties of FR~0s}

The spectral index information can be used to probe the presence of
optically-thin emission in FR~0s, and also on scales smaller than the
spatial resolution of the TGSS (10 - 25 kpc).  The fraction of
\FRo\ sources detected by the TGSS is 42\%: this relatively low
detection rate is due to the combination of (1) the different flux
density thresholds used to build the \FRo\ catalog and the one adopted for the
TGSS (5 and 17.5 mJy, respectively), and, (2) the generally flat spectral
slope between 150 MHz  and 1.4 GHz of the FR~0s.  Nonetheless, the TGSS
sensitivity is sufficient to exclude that the sources not detected by
this survey have a spectrum with a slope steeper than 0.5. This leaves
us with only eight FR~0s with a steep low-frequency radio spectrum.

A simple model in which the emission is produced by two components,
one extended and one compact, with spectral index $\alpha$=0.7 and
$\alpha$=0, respectively, indicates that the overall spectrum becomes
steep ($\alpha > 0.5$) when the optically-thin component contributes
to a fraction $>$55\% of the total emission at 1.4 GHz.  Therefore, at
least half the emission from the 35 flat or inverted sources must
originate from a core component. This value confirms the high-core
dominance of FR~0s derived by \citet{baldi19}, based on high resolution
images.

\subsection{Contribution of young radio-galaxies to the FR~0s' population}

A further issue that can be investigated from the low-frequency
spectral properties of FR~0s is how many of them are compact because
of their youth. Although, as reported in the introduction, the number
density of FR~0s is too large to interpret the whole class of compact
sources as young objects, the sources that will eventually produce
extended RGs must necessarily pass through a small-size phase, and some
FR~0s might indeed compact because they are young.

Young RGs can be found looking for a convex spectrum due to a low
frequency cut-off.  By restricting to the 19 FR~0s with $F_{1.4}
  > 50$ mJy, the fraction of sources with a convex spectrum is $\sim$
75\%.  Nonetheless, the spectral curvature of the FR~0s spectra is
less pronounced than what is measured in the more powerful GPS
sources. \citeauthor{odea91} fit the radio spectra of a sample of GPS
with log-parabola (the same method we adopted in the previous section)
and found that generally the peak of their radio spectra are rather
narrow, with a median value of the FWHM of 1.2 decades of
frequency. Conversely, we find larger values, typically between
  1.5 and 6 decades (with a median value of $\sim$2.3). Although five
  of them must be considered as upper limits, because the source is
not detected in at least one band, the spectral curvature of FR~0s is
generally much less pronounced than in GPS.

Apparently, we do not see the sharp transition from an optically-thick to an optically-thin regime typical of the GPS. The spectral
properties of FR~0s are better described as being due to a gradual
steepening toward high frequencies.  \citet{baldi19} noted a similar
effect in their VLA observations between 1.4 and 7.5 GHz. The
  spectrum between 4.5 and 7.5 GHz is steeper than between 1.4 and 4.5
  GHz, meaning, the high frequency spectra of FR~0s are generally
convex: the median difference in the spectral slopes from 1.4 to 5 GHz
and from 4.5 to 7.5 GHz is $\Delta \alpha =
0.16$. \citeauthor{baldi19} also found six out of 18 sources have
  steep spectra (the remaining objects have flat and, in one case,
inverted, spectra), a significantly higher fraction than what we find
in this study. Interestingly, the steep sources in the
sub-sample\footnote{J0907+35, one of the six belonging to this group,
  is the only FR~0s not covered by the TGSS.} have a flat spectrum
between 150 MHz and 1.4 GHz, confirming the presence of a spectral
steepening, with data covering a broader spectral range.

The presence of sources in which the spectral index increases with
frequency, similarly to what we see in FR~0s, has been already
recognized in very early works and ascribed to the combined effects of
self-absorption and of the presence of components with a wide range of
brightness temperatures \citep{kellermann69,marscher88}. This
interpretation might also apply to FR~0s, and it can be tested with
radio imaging at a high resolution, sufficient to spatially resolve the
various emitting components.

Nonetheless, in the bright sub-sample of 19 FR~0s, there are three
sources (namely J0906+41, J1116+29, and J1250+00) whose spectrum is
reminiscent of the GPS spectra: they have an inverted spectrum at
low-frequencies, and an emission peak at $\nu \gtrsim 1$GHz. The FR~0s
in which we might be observing a genuine low-frequency cut-off, and
which can be interpreted as young compact objects, represent $\sim
15\%$ of the sub-sample with flux densities larger than 50
mJy. However, this result must be confirmed with more sensitive
surveys. In fact, this flux density limit at 1.4 GHz might introduce a
significant bias, because the median luminosity of this sub-sample is a
factor 10 higher than for the whole \FRo. Furthermore, we might be
excluding sources in which the emission peak is located at higher
frequencies, and which, for this reason, do not meet the flux density
threshold. The very selection of the \FRo\ is based on surveys at 1.4
GHz, and this might represent a bias against sources with a GPS-like
spectra.

Furthermore, variability is known to play an important role in the
process of identification of this class of sources (see, e.g.,
\citealt{torniainen05}): simultaneous multi-frequency observations are
needed to firmly assess their nature. The study of these candidates is
particularly relevant because of their extremely low radio luminosity
($2 - 3 \times 10^{23} \WHz$), more than five orders of
magnitude below that of the most studied samples of young compact
sources \citep{odea98}, and even less powerful than the low-luminosity
compact sources studied by \citet{kunert10}.

\section{Summary and conclusions}
\label{summary}

We present the results obtained from the TGSS survey, based on
  GMRT observations at 150 MHz of the 104 compact FR~0s sources
  forming the \FRo\ sample. The fraction of FR~0s detected at low
  radio frequencies is 36\%. The relatively small number of 150 MHz
  detections is due in part to the higher flux density threshold of the
  TGSS with respect to the selection threshold of \FRo\ (17.5 and 5
  mJy, respectively), but also to the general flatness of the radio
  spectra: only eight sources have a steep ($\alpha > 0.5$) radio
  spectral index between 150 MHz and 1.4 GHz.
  
We failed to detect extended emission associated with the FR~0s. The
corresponding upper limit, estimated over a region of 100 kpc $\times$
100 kpc, is 4$\times 10^{23}$ W Hz$^{-1}$, a factor between 3 and 300
below the luminosity of FR~I sources. In addition, the FR~0s' spectral
shapes indicate that the contribution of extended optically-thin
emission within the TGSS beam might contribute for, on average, at
most a fraction $\lesssim$ 50\% (at 1.4 GHz) to these compact sources.

By also including observations at 5 GHz from the GB6 survey, we
explored the radio spectra of FR~0s over a larger range of
frequencies. Due to the higher threshold of the GB6 ($\sim$ 18 mJy), only
23 FR~0s are detected at 5 GHz. We then preferred to limit the
multi-band analysis to the sub-sample formed by the 19 FR~0s with a
flux density at 1.4 GHz larger than 50 mJy. Most of them (13) have
spectral indices flatter than 0.5 in both frequency ranges, and in 14
FR~0s, the spectra is steeper between 1.4 and 5 GHz than between 150
MHz and 1.4 GHz, meaning they are convex spectra.

A convex spectrum is a characteristic feature of young sources in
which a turn-over is observed at low frequencies, due to a high
optical depth of either free-free or synchrotron self-absorption. This
raised the possibility that at least a fraction of the \FRo\ sources
are compact, because they are young radio-galaxies. Nonetheless the
spectral curvature of FR~0s is in general smaller than in GPS: in
FR~0s, the median FWHM is 2.3 decades of frequency compared to a FWHM
of 1.2 measured in GPS. The fraction of FR~0s with a high curvature
and a spectrum rising in the GHz spectral region, reminiscent of GPS,
is at most three out of 19, meaning $\lesssim$ 15\%.

Clearly, the studies of the low-frequency radio properties of FR~0s
would greatly benefit from the deeper and higher resolution
observations that are being produced by the International Low
Frequency Array (LOFAR; \citealt{vanhaarlem13}).  In particular, the
LOFAR Two-meter Sky Survey (LoTSS) will eventually cover the entire
Northern sky, producing $\sim$5\arcsec\ resolution images with a
sensitivity of $\sim$ 0.1 mJy beam$^{-1}$ at 150 MHz
\citep{shimwell17}. It should enable us to set stronger limits on, or
even allow the detection of, the extended emission, with an
improvement of the detection threshold of more than an order of
magnitude with respect to the TGSS, and to detect the low-frequency
counterpart of all the FR~0s with $\alpha(150-1400) > -1$. This will
enable us to characterize the spectral shape of
FR~0s in much greater detail, in particular for those with a convex spectrum.

\begin{acknowledgements}
MB acknowledges support from INAF under PRIN SKA/CTA FORECaST and from
the ERC-Stg DRANOEL, no 714245.  RDB has received funding from the
European Union’s Horizon 2020 research and innovation
programme under grant agreement No 730562 [RadioNet]. TOPCAT
astronomical software \citep{taylor05} was used for the preparation
and manipulation of the tabular data and the images.  We thank the
staff of the GMRT that made these observations possible. GMRT is run
by the National Centre for Radio Astrophysics of the Tata Institute of
Fundamental Research. This research has made use of Aladin sky atlas
developed at CDS, Strasbourg Observatory, France.
\end{acknowledgements}
  
\bibliographystyle{aa}

\end{document}